\documentclass[final]{cvpr}
\usepackage{algorithmic}
\usepackage{graphicx}
\usepackage{textcomp}
\usepackage{times}
\usepackage{epsfig}
\usepackage{graphicx}
\usepackage{amsmath}
\usepackage{amssymb}

\usepackage{multirow}

\usepackage[pagebackref=true,breaklinks=true,colorlinks,bookmarks=false]{hyperref}

\begin{document}

\title{De-Noising of Photoacoustic Microscopy Images by Deep Learning}

\author{Da He\textsuperscript{1}\thanks{Contributed equally.}, Jiasheng Zhou\textsuperscript{1}\footnotemark[1], Xiaoyu Shang\textsuperscript{1}, Jiajia Luo\textsuperscript{2}\thanks{Corresponding authors.}, and Sung-Liang Chen\textsuperscript{1}\footnotemark[2]\\
\textsuperscript{1}University of Michigan-Shanghai Jiao Tong University Joint Institute,\\Shanghai Jiao Tong University, Shanghai 200240, China\\
\textsuperscript{2}Biomedical Engineering Department, Peking University, Beijing 100191, China\\
{\tt\small \{da.he, jiasheng\_zhou, shangxiaoyu, sungliang.chen\}@sjtu.edu.cn, jiajia.luo@pku.edu.cn}
}
\maketitle

\begin{abstract}
As a hybrid imaging technology, photoacoustic microscopy (PAM) imaging suffers from noise due to the maximum permissible exposure of laser intensity, attenuation of ultrasound in the tissue, and the inherent noise of the transducer. De-noising is a post-processing method to reduce noise, and PAM image quality can be recovered. However, previous de-noising techniques usually heavily rely on mathematical priors as well as manually selected parameters, resulting in unsatisfactory and slow de-noising performance for different noisy images, which greatly hinders practical and clinical applications. In this work, we propose a deep learning-based method to remove complex noise from PAM images without mathematical priors and manual selection of settings for different input images. An attention enhanced generative adversarial network is used to extract image features and remove various noises. The proposed method is demonstrated on both synthetic and real datasets, including phantom (leaf veins) and \emph{in vivo} (mouse ear blood vessels and zebrafish pigment) experiments. The results show that compared with previous PAM de-noising methods, our method exhibits good performance in recovering images qualitatively and quantitatively. In addition, the de-noising speed of 0.016 s is achieved for an image with $256\times256$ pixels. Our approach is effective and practical for the de-noising of PAM images. 

\noindent
{\bf Key words:}

Photoacoustic microscopy, de-nosing, generative adversarial network, deep learning.
\end{abstract}

\section{Introduction}
\label{sec:introduction}
Photoacoustic (PA) imaging (PAI), based on the PA effect, is a new non-invasive imaging technology, which images the object according to the PA signal generated upon laser illumination of the object~\cite{a1,a5}. The advantage of PAI is that it combines the highlights of optical and ultrasonic imaging with good performance in terms of contrast and penetration. PAI can be implemented as PA computed tomography (PACT) and PA microscopy (PAM). The former forms images by image reconstruction algorithms, while the latter performs raster scanning without complicated image reconstruction~\cite{a5}. PAI has potential applications in vascular biology, ophthalmology, dermatology, neurology, etc.~\cite{a1}.

High signal-to-noise ratio (SNR) is the key to high-quality PA images, which are essential for \emph{in vivo} animal studies and clinical applications. A pulsed laser with pulse duration of a few nanoseconds is typically used for efficient PA conversion. The maximum PA signal amplitude is in part limited by allowable laser fluence on the tissue surface and is capped by the ANSI safety limit. PA contrast agents, as exogenous absorbers, are a common solution to enhance the PA signal amplitude~\cite{a7}. Development of PA contrast agents for \emph{in vivo} and clinical applications usually requires cumbersome effort. On the other hand, in PA signal and image acquisition, noise is generated from laser illumination to signal detection~\cite{a9,a10,a11}. The PA noise in PA signals and images arises from several different factors, and building an accurate PA noise model can be complicated. Overall, the PA noise can be categorized into source-related noise (e.g., fluctuations in laser illumination) and system-related noise (e.g., randomly distributed thermal and electronic noise)~\cite{a10}. Among the different types of thermal noise, white Gaussian noise is the most common one~\cite{a10}. Besides boosting the PA signal amplitude with contrast agents, de-noising is an alternative and promising approach to enhance SNR.

Several PA de-noising methods have been demonstrated to enhance the SNR of the PA signals and/or images~\cite{a12,a18,a19,a20,a21,a22,a24,a26,a27,a28,a30}. The de-noising has been studied for PACT~\cite{a12,a18,a19,a20,a22,a28,a30} and PAM~\cite{a21,a26,a27}. For most PACT de-noising, PA raw data (A-line signals) are usually de-noised first prior to image reconstruction~\cite{a12,a19,a22}. For PAM, de-noising may be conducted for PA A-line signals~\cite{a26,a27} or for PA images directly~\cite{a21,a24}. De-noising of \emph{in vivo} images was also tested in some of these works~\cite{a18,a19,a20,a21,a24,a26,a27,a30}. Among these PA de-noising techniques, signal averaging is the most commonly used one~\cite{a19}, yet acquisition of multiple signals is needed, which is time-consuming and results in low temporal resolution. Further, researchers have made effort to enhance SNR using various methods to handle the PA raw A-line signals, such as wavelet-based algorithms~\cite{a12,a22}, empirical mode decomposition (EMD)~\cite{a26,a27}, and sparsity-based methods~\cite{a18,a28}. Alternatively, methods applied in the image domain, such as K-means singular value decomposition (KSVD) and non-local means (NLM)~\cite{a21,a24}, can be used to de-noise PA images directly. However, these methods suffer one or some of the issues: (i) Prior information about the noise property is needed, which is a significant challenge~\cite{a30}; (ii) time-consuming computation is required, especially for those based on iterative optimization; (iii) some parameters such as the noise level are needed to be manually specified for different input signals and images. 

In recent years, deep learning based on convolutional neural networks (CNNs) has been extensively used for medical imaging~\cite{a31}. There are many applications to medical imaging such as classification between lesions and non-lesions, classification of lesion types, detection of lesions, etc.~\cite{a31}. Deep learning has also been used in PAI, such as reflection artifact removal in PACT images, PACT image reconstruction with sparse data, and PAM imaging with sparse data~\cite{a32,a33,a34}. Besides, deep learning has shown promise in de-noising natural images~\cite{a38,a40} as well as medical images including computed tomography (CT)~\cite{a42}, ultrasound imaging (de-speckling)~\cite{a43}, and optical coherence tomography~\cite{a44}. Especially, generative adversarial network (GAN), which utilizes a discriminator network to guide the distribution of generated samples of a generator network~\cite{a46}, was used and showed great performance in de-noising~\cite{a38,a40,a42,a43,a44}. Recently, PA de-noising based on deep learning has been investigated~\cite{a30,a}. Although Hariri et al. applied CNN to PA de-noising~\cite{a30}, PACT images with relatively low resolution and sparse patterns (in contrast to PAM images) were studied. Sharma et al. studied resolution enhancement in AR-PAM images using CNN~\cite{a}. In that work, a CNN architecture was developed mainly to improve out-of-focus lateral resolution of AR-PAM images, while background noise reduction was also observed. Zhao et al. demonstrated good de-noising performance for OR-PAM images~\cite{x2}, yet the use of two wavelengths for PA excitation was required, which results in higher cost and more complexity of the imaging system (e.g., an optical parametric oscillator laser was used.). 

In this work, we aim to de-noise PAM images based on the deep learning method. As mentioned above, GAN has been demonstrated to be powerful in de-noising natural and medical images~\cite{a38,a40,a42,a43,a44}, while GAN-based PA de-noising work is still limited. Inspired by these works, we propose a GAN-based CNN to de-noise PAM images with the following key contributions. First, we develop a CNN network with key features including: (i) Attention modules in the generator, which effectively captures the pattern relationships in 2D images and thus can distinguish signal pixels from background noise; (ii) a novel composite loss function, which excels in restoring fine features and thus enables faithful recovery from noisy images. By contrast, a pixel-based loss function tends to ignore high-frequency details. Secondly, we demonstrate that effective de-noising of PAM images using the developed CNN network outperforms the existing methods using block-matching and 3D filtering (BM3D)~\cite{a47}, KSVD~\cite{a21}, and weighted nuclear norm minimization (WNNM)~\cite{a48}. This is evidenced by quantitative and qualitative comparisons. Finally, we explore how well our CNN network (trained with only the synthetic dataset) de-noises real experimental datasets of mouse ear vasculature and zebrafish pigment. The results show that our CNN network is promising for \emph{in vivo} PAM de-noising applications.

\section{Method}
%

The proposed GAN-based CNN model consists of three main components: a generator network, a discriminator network, and a perceptual loss calculator (VGG network). Let $I_N$, $I_C$, and $\widehat{I_C}$ ($I_N$, $I_C$, $\widehat{I_C} \in R^{M\times N}$) be the noisy PA image, the corresponding real clean image (if any), and the corresponding de-noised image, respectively, where $M$ and $N$ denote the height and width of the image, respectively. Then, our objective is to train a generator $G$ that maps $I_N$ to $\widehat{I_C}$ ($G:I_N \rightarrow \widehat{I_C}$). Meanwhile, the discriminator D tries to distinguish the real clean PA image ($I_C$) from the de-noised PA image ($\widehat{I_C}$). The training of the generator network against the discriminator forms the adversarial min-max problem~\cite{a46}.

\begin{figure*}[ht]
	\begin{center}
		\includegraphics[width=1\linewidth]{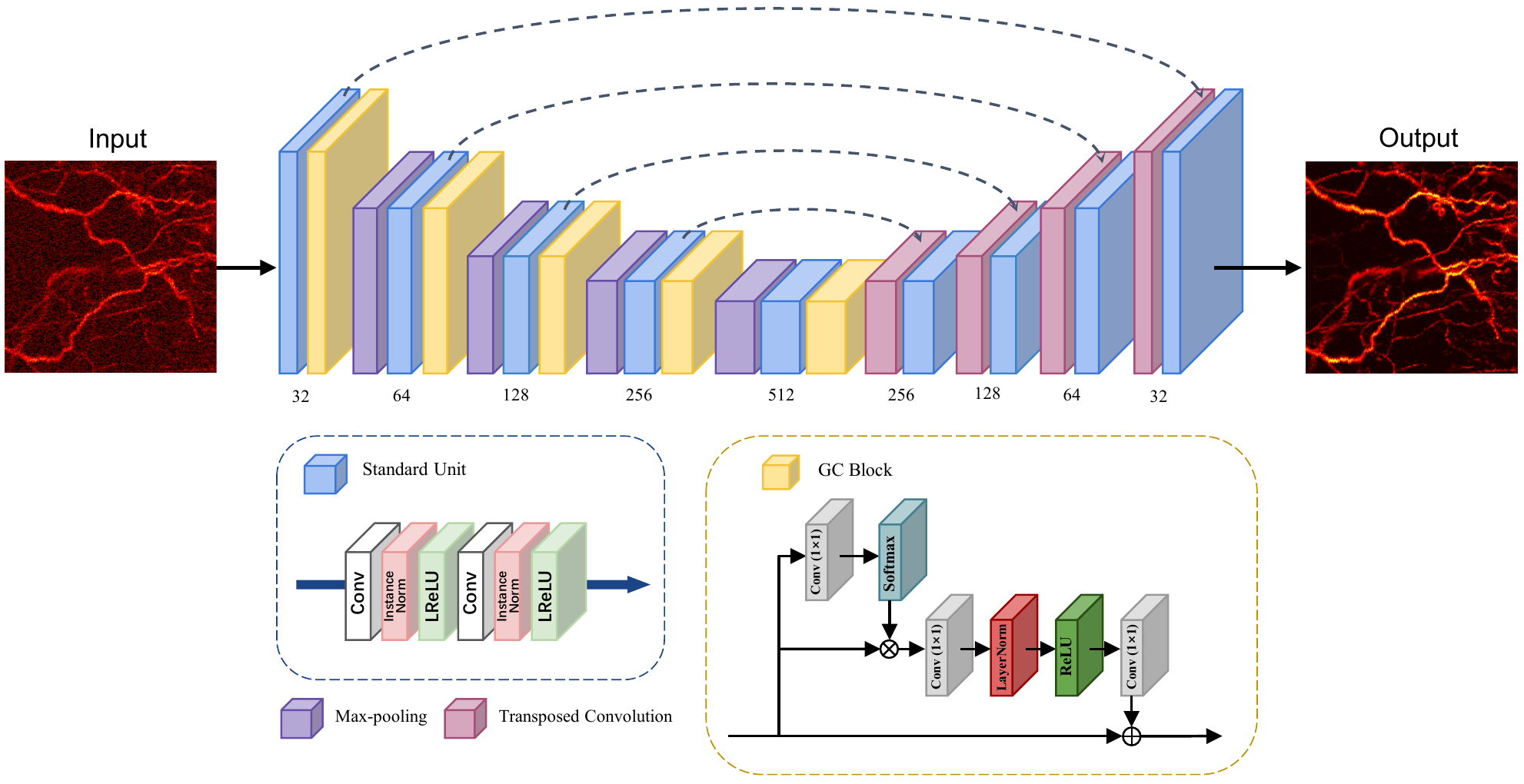}
	\end{center}
	\caption{Architecture of the proposed generator. The number of channels for the output tensor is indicated below the corresponding layer.}
	\label{fig:1}
\end{figure*}

For the generator network, we incorporate a global context (GC) attention module into a modified U-Net style structure to generate the de-noised image as output. In order to optimize the generator network, we elaborate a combined loss function comprising the adversarial loss, the perceptual loss, and the pixel-level loss. 

In the following subsections, first, the noise model used to produce synthetic noisy PA images is discussed. Secondly, the architecture of our CNN model is detailed, including the attention module and the combined loss function. Thirdly, the dataset, network implementation, and evaluation metrics in this study are presented. Finally, other de-noising methods for comparison with our proposed CNN-based method are introduced.

\subsection{PA Image Noise Model}
\label{sec:noise_model}

Typically, noise in the PAI process can be simply modeled as the combination of Gaussian noise, Poisson noise, and Rayleigh noise. Gaussian noise is a basic noise model to account for a few types of noise including thermal, amplifier, and read noise. Gaussian noise is fully independent of signals for a particular experimental system and thus can be added (i.e., additive noise) to any other noise that might be intrinsic to the system. In PA de-noising works, Gaussian noise is commonly assumed and modeled~\cite{a18,a21,a28}. On the other hand, the signal-dependent part of the noise can be modeled as Poisson noise, which can be introduced in the process of signal conversion and transmission relating to the fluctuation and digitalization of particles in electronic devices~\cite{b}. Rayleigh noise may be introduced by reflection and scattering of PA waves in tissue transmission~\cite{c}.

In real experimental PA images, the noise is usually complicated and non-uniformly distributed over the entire image. Therefore, there is no clear clue how the noisy PA images and their latent clean PA images are interrelated, which makes it difficult to de-noise PA images using traditional methods. Alternatively, one can use a CNN model trained on a noisy PA image dataset, e.g., using a synthetic dataset through introducing the noise into clean PA images, to de-noise real noisy PA images.

\subsection{Network Architecture}
\subsubsection{Generator Network}

The generator framework (Fig.~\ref{fig:1}) applies the U-Net shape architecture and contains layers of pooling and transposed convolution, standard unit blocks, and GC blocks~\cite{a50}. U-Net shape networks have shown high performance in medical image processing due to multi-scale feature fusion and lightweight parameters. The proposed standard unit block consists of two $3\times 3$ convolutional layers, each followed by an instance normalization layer~\cite{a55} and a leaky ReLU (LReLU) function. All nine standard unit blocks in the generator have 32, 64, 128, 256, 512, 256, 128, 64, and 32 filters, respectively. 

To efficiently capture features and distinguish information with varying importance, an attention mechanism is applied in our network. Unlike regular CNNs that may treat all information equally, attention blocks additionally introduce attention weights for different feature channels or spatial positions. Specifically, the proposed method utilizes the attention block, i.e., the GC block, to enhance the attention to long-range dependencies and thus better handle unexpected noise instead of focusing on signal pixels. The detailed structure of the GC attention block includes $1\times1$ convolutions and layer normalizations as described in Fig.~\ref{fig:1}. In the generator, GC blocks are placed after each standard unit block of the encoder.
\begin{figure}[th]
	\begin{center}
		\includegraphics[width=1\linewidth]{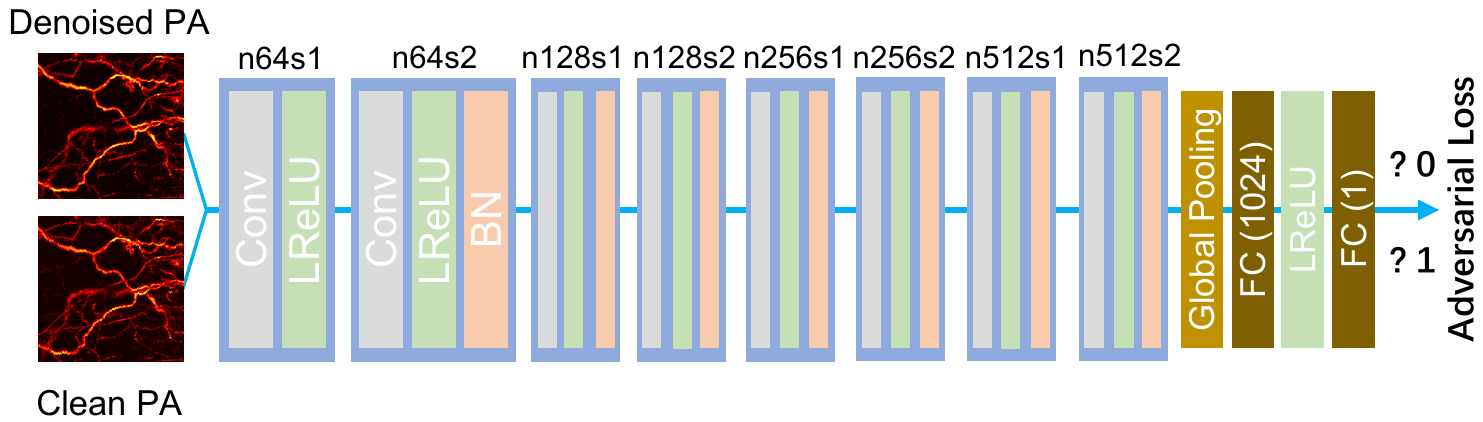}
	\end{center}
	\caption{Architecture of the discriminator.}
	\label{fig:2}
\end{figure}
Then, in the decoding process, the generator performs direct up-sampling of the previously extracted high-level feature maps, combining global and local features, until the size (in pixels) of the original image is restored. The max-pooling layers and the transposed convolutional layers are used for down-sampling and up-sampling in the encoder and decoder parts, respectively. The generator takes the noisy image as input and outputs its de-noised version. 

\subsubsection{Discriminator Network}

The discriminator illustrated in Fig.~\ref{fig:2} is an 8-layer CNN with the number of filters as 64, 64, 128, 128, 256, 256, 512, and 512, respectively . Each of the convolutional layers has a kernel size of $3\times3$ and is subsequently equipped with a LReLU and a batch normalization (BN). In the end, there are two fully-connected (FC) layers with 1024 outputs and a single output, respectively. The input of the discriminator is the de-noised image or its corresponding clean image (ground truth), and the output is the probability that the de-noised image is true.

\subsection{Combined Loss Function}
The loss function for our model consists of three parts: (i) perceptual loss~\cite{a56} $L_{perceptual}$, (ii) smooth L1 loss~\cite{a57} $L_{smoothL1}$, and (iii) adversarial loss $L_{GAN}$. It is expressed as:
\begin{equation}
L = k_1 L_{perceptual} + k_2 L_{smoothL1} + k_3 L_{GAN},
\end{equation}
where $k_1$, $k_2$ and $k_3$ are hyperparameters to control the trade-off among all components.

\subsubsection{Perceptual Loss}
Most de-noising algorithms, including deep learning-based methods, aim to minimize the mean square error (MSE) between the de-noised image ($\widehat{I_C}$) and the ground truth ($I_C$). However, using only MSE loss may produce blurred images with loss of details and faithfulness~\cite{a34}. To better restore images, we adopt the perceptual loss, which focuses on the high-level features to make the restoration performance more consistent with the perception of the human visual system. Specifically, the high-level features are extracted from the pre-trained VGG-19 model~\cite{a58}. The perceptual loss can be formulated as follows:
\begin{equation}
\begin{aligned}
&L_{perceptual}(\widehat{I_C}, I_C)\\
& = \frac{1}{HWC}\sum_{i,j,k}(VGG_{i,j,k}(\widehat{I_C}) - VGG_{i,j,k}(I_C))^2,
\end{aligned}
\end{equation}
where $H$, $W$, and $C$ are the height, width, and channel size of the input tensor.

\subsubsection{Smooth L1 Loss}
Smooth L1 loss can avoid the defects of the regular L1 loss and L2 loss, alleviating the gradient explosion of L2 loss for large errors and also improving the stability of L1 loss for small errors. Therefore, the applied smooth L1 loss is expressed as follows:
\begin{equation}
L_{smoothL1}(\widehat{I_C}, I_C) = \frac{1}{HWC}\sum_{i,j,k}\mathcal{L}(\widehat{I_C}_{i,j,k} - {I_C}_{i,j,k}),
\end{equation}
\begin{equation}
\mathcal{L}=\left\{
\begin{aligned}
&0.5(\hat{x} - x)^2 \text{, if }|\hat{x}-x|<1\\
&|\hat{x}-x|-0.5 \text{, otherwise}
\end{aligned}
\right.
\end{equation}
where $|\hat{x}-x|$ is the pixel difference between the restored image patch and the clean one.

\subsubsection{GAN Loss}
A generative component is added to our GAN to force the restored image to be as realistic as the clean image, so that the restored image can deceive the discriminator. The adversarial loss over $BS$ training samples is expressed as follows:
\begin{equation}
L_{GAN}(G)=\sum^{BS}[-\log D(G(I_N))],
\end{equation}
where $D(G(I_N))$ is the estimated probability that the de-noised image $G(I_N)$ (i.e., $\widehat{I_C}$) is a real clean PA image.

\subsection{Dataset}
\label{sec:dataset}
Supervised de-noising training is expected to receive numerous image pairs with corresponding noisy and clean images; however, obtaining paired biomedical PAM images in experiments is difficult and time-consuming. To handle the limitation, a synthesis operation is applied to construct the training dataset. Since the training dataset employs an inclusive noise model that considers three types of noise, the CNN model trained from the dataset is expected to perform well on synthetic datasets as well as real testing datasets.

We first introduce the preparation of clean PAM images of leaf veins and mouse ear blood vessels as the clean images used in this work. Then, we describe the dataset of synthetic noisy images used for training. Finally, we present the acquisition of the dataset of real noisy images. The details about the PAM experimental setup and image acquisition can be found in our recent paper~\cite{a34}. Note that in PAM images, 2D maximum amplitude projection (MAP) images (projection along the axial direction) are typically used. Therefore, in this work, we aim to restore PAM MAP images with good de-noising effect. The animal experiment was conducted in conformity with the laboratory animal protocol approved by Laboratory Animal Care Committee of Shanghai Jiao Tong University.

\subsubsection{Clean Leaf Vein Dataset}
We adopted the dataset of PAM images of bodhi and magnolia leaf veins acquired in our previous work~\cite{a34}. The size of the raw images is $256\times256\times180$, where 180 is the number of pixels along the axial direction. Thus, the size of the 2D MAP images is $256\times256$. In total, there are 260 leaf vein PAM images as the clean leaf vein dataset used in this work.

\subsubsection{Clean Blood Vessel Dataset}
\label{sec:clean_vessel_dataset}
We experimentally acquired many PAM images of mouse ear blood vessels \emph{in vivo}. In total, 165 image patches with size of $250\times250\times180$ were prepared as the clean blood vessel dataset. Compared with PAM images of leaf veins, those of blood vessels have relatively rich details and complex structure. Thus, the de-noising of PAM images of blood vessels is expected to be more challenging.

\subsubsection{Synthetic Noisy Image Dataset}
The noise model described in subsection~\ref{sec:noise_model} can be used to synthesize noisy images. To better mimic the noise generation during PAM image acquisition, instead of adding noise directly to the 2D MAP images, noise was added to PA A-line signals, and then MAP was applied to obtain noisy 2D images. All three types of noise (i.e., Gaussian, Poisson, and Rayleigh noise) were randomly generated by Python functions, with noise levels ranging from mild to severe. For the clean leaf vein dataset, 236 images were used to construct clean-noisy image pairs for training, and 24 images were used for testing. As for the clean blood vessel dataset, similarly, 149 images were used for training and 16 images for testing. 

\subsubsection{Real Noisy Image Dataset}
\label{sec:real_noisy_dataset}
In PAM imaging, the higher the excitation light energy used, the better the SNR obtained. Therefore, a real noisy PAM image dataset with low SNR was obtained by using relatively low excitation light energy. In total, we acquired 16 real noisy images of leaf veins with size of $256\times256$ and 21 real noisy images of blood vessels with size of $250\times250$ to test the performance of our CNN model. Further, to test the generalization ability of our CNN model, we also used real noisy images of zebrafish pigment acquired in our previous work~\cite{x1}.

\subsection{Network Implementation}
To evaluate the performance of our CNN model, we trained two models using different synthetic noisy image datasets (including leaf veins and mouse ear blood vessels, as mentioned in III.D). We adopted Adam optimizer with parameters $\beta_1=0.9$, $\beta_2=0.999$, and $\epsilon=10^{-8}$~\cite{a60}. We set the learning rate as 0.0001, the batch size as 8, and the epoch number as 60,000. During training, the $k_2$ in the combined loss was initially 1.0 and finally decreased to 0, while $k_1$ and $k_3$ increased from 0 to 1.0 and $10^{-3}$, respectively. All the models were implemented using TensorFlow (v. 1.13.1) based on Python (v. 3.5.2). The training of each model took about 8 hours on an NVIDIA TITAN RTX GPU with 24 GB memory.

\subsection{Evaluation Metrics}
\subsubsection{Full-reference Metrics}
For the synthetic noisy image dataset with ground truth, to quantitatively evaluate de-noising performance, we used the metrics of peak SNR (PSNR) and structural similarity (SSIM).

\begin{figure}[b]
	\begin{center}
		\includegraphics[width=0.6\linewidth]{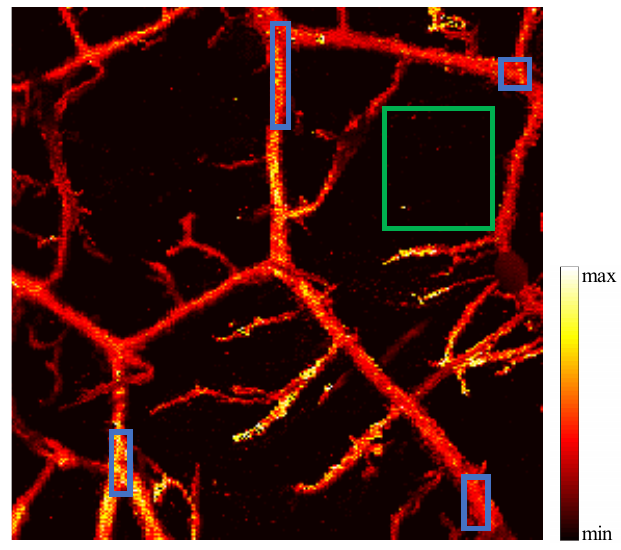}
	\end{center}
	\caption{Illustration of the signal ROIs (the blue boxes) and background ROI (the green box).}
	\label{fig:3}
\end{figure}

\begin{table*}[t]
	\caption{Quantitative Comparison Results Using the Synthetic Leaf Vein Dataset (Mean$\pm$Standard Deviation)}
	\small
	\begin{center}
		\begin{tabular}{c|ccccc}
			\hline
			Method & Noisy input & BM3D & KSVD & WNNM & Ours \\
			\hline
			PSNR (dB) & 21.79$\pm$4.08 & 21.97$\pm$3.91 & 21.88$\pm$3.93 & 22.15$\pm$4.04 & \bf{26.81$\pm$1.93} \\
			\hline
			SSIM & 0.34$\pm$0.14 & 0.34$\pm$0.14 & 0.34$\pm$0.14 & 0.35$\pm$0.14 & \bf{0.84$\pm$0.08} \\
			\hline
		\end{tabular}
	\end{center}
	\label{table:1}
\end{table*}

\begin{figure*}[ht]
	\begin{center}
		\includegraphics[width=1\linewidth]{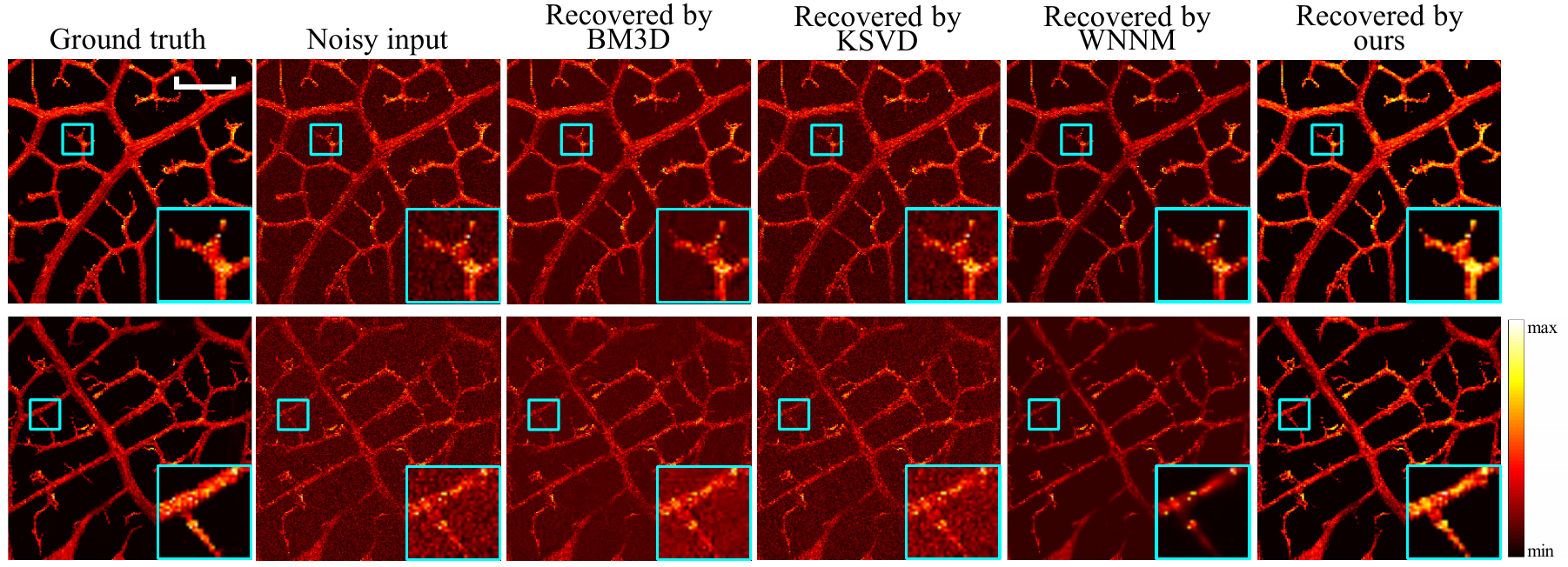}
	\end{center}
	\caption{Representative results from the testing set of the synthetic leaf vein dataset. The top row is from a bodhi leaf and the bottom row comes from a magnolia leaf. Scale bar: 500 \textmu m. All images, excluding zoom images, share the same scale bar. The values in the colorbar indicate relative PA intensity.}
	\label{fig:4}
\end{figure*}

\subsection{No-reference Metrics}
By contrast, for the real noisy image dataset without ground truth, the de-noising performance was evaluated by the no-reference metrics of SNR and contrast-to-noise ratio (CNR). To calculate these metrics, representative signal regions and background regions from each sample should be selected as the regions of interest (ROIs). SNR and CNR can be calculated as follows, respectively~\cite{a10}:
\begin{equation}
SNR = 20\log\left(\frac{\frac{1}{n}\sum_{i=1}^{n}\mu_i}{\sigma_b}\right),
\end{equation}
\begin{equation}
CNR = 20\log\left(\frac{1}{n}\sum_{i=1}^{n}\frac{|\mu_i - \mu_b|}{\sqrt{\sigma_{i}^{2} + \sigma_{b}^{2}}}\right),
\end{equation}
where $\mu_i$ and $\sigma_i$ denote the mean and standard deviation of the ith signal ROI, respectively; $\mu_b$ and $\sigma_{b}$ are the mean and standard deviation of the background ROI; $n$ is the number of signal ROIs, which was set as 4 in our evaluation. The signal and background ROIs are illustrated as blue and green boxes, respectively, as shown in Fig.~\ref{fig:3}.

\subsection{Comparison with Other Methods}
In this work, we compared our proposed method with three common de-noising algorithms, which are based on an image non-local self-similarity (NSS) model named BM3D~\cite{a47}, a sparse model named KSVD~\cite{a61}, and a low-rank model named WNNM~\cite{a48}.

\section{Results}
\label{sec:results}
\subsection{Synthetic Leaf Vein Dataset}
\label{sec:synthetic_results}
As described previously, 260 leaf vein PAM images were used to implement the de-noising experiment. These clean PAM images were regarded as the ground truth while their corresponding synthetic noisy images were used as input, forming a supervised learning manner to train the proposed de-noising network. Specifically, after randomly adding three kinds of noise, there are 808 image pairs for training and 83 image pairs for testing. After training, the de-noising model was evaluated using the testing set. For fair comparison, all the following quantitative and qualitative results were generated based on the testing set only.

Because the synthetic experiment includes both the noisy input and the ground truth of PAM images, full-reference metrics of PSNR and SSIM can be used to provide a good measure of the de-noising performance (i.e., the similarity between the de-noised result and the ground truth). Table I shows the statistical results on the testing set.

\begin{table*}[ht]
	\caption{Quantitative Comparison Results Using the Real Noisy Leaf Vein Dataset (Mean$\pm$Standard Deviation)}
	\small
	\begin{center}
		\begin{tabular}{c|ccccc}
			\hline
			Method & Noisy input & BM3D & KSVD & WNNM & Ours \\
			\hline
			SNR (dB) & 29.08$\pm$7.96 & 36.23$\pm$5.36 & 29.20$\pm$6.58 & 55.53$\pm$15.65 & \bf{90.73$\pm$15.86} \\
			\hline
			CNR (dB) & 4.80$\pm$2.59 & 5.37$\pm$2.09 & 5.53$\pm$2.46 & 5.72$\pm$2.25 & \bf{7.63$\pm$1.77} \\
			\hline
		\end{tabular}
	\end{center}
	\label{table:2}
\end{table*}

\begin{figure*}[ht]
	\begin{center}
		\includegraphics[width=1\linewidth]{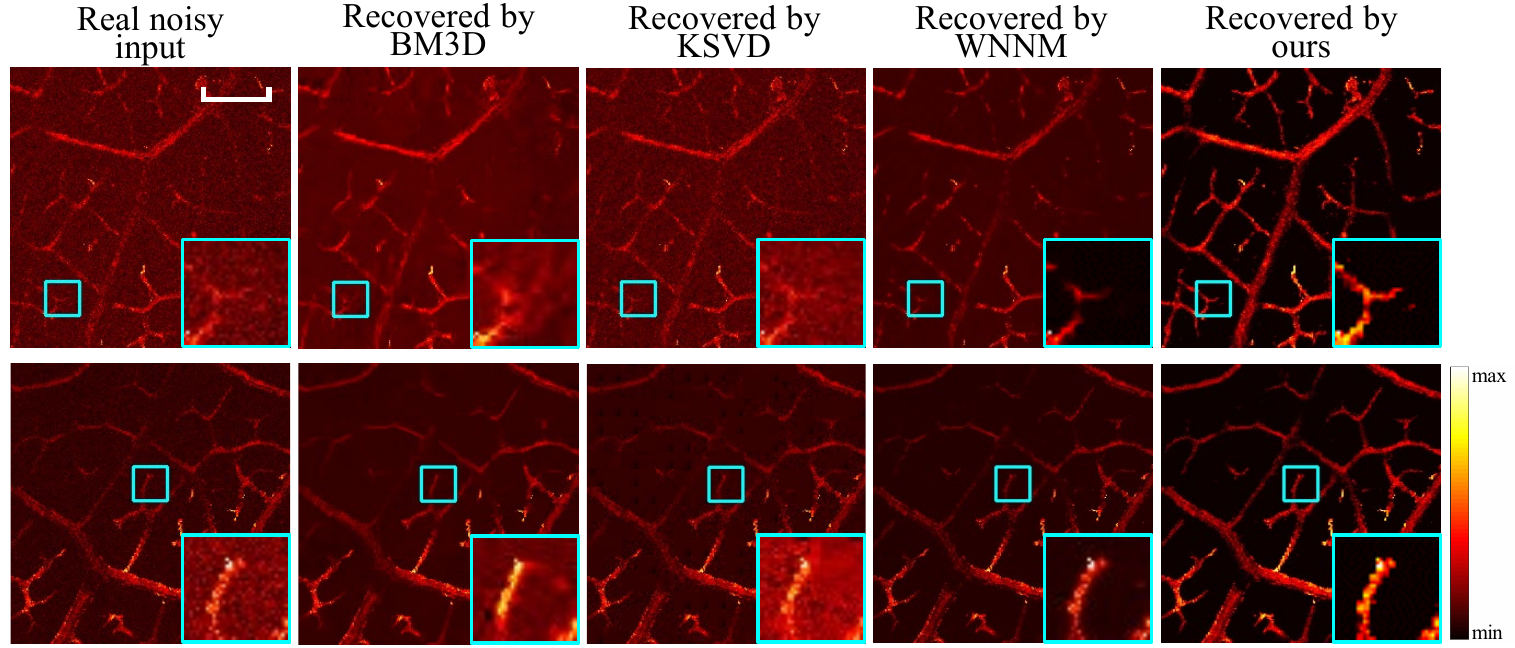}
	\end{center}
	\caption{Representative results from the experimentally-acquired noisy leaf vein dataset. Both rows are from bodhi leaves. Scale bar: 500 \textmu m. All images, excluding zoom images, share the same scale bar. The values in the colorbar indicate relative PA intensity.}
	\label{fig:5}
\end{figure*}

According to Table~\ref{table:1}, our method outperforms other methods with the obvious margin in terms of both PSNR and SSIM. After adding various types and levels of noise, the noisy images in the testing set have the average PSNR and SSIM of 21.79 dB and 0.34, respectively, which indicate that the image quality is quite low. Compared to the noisy input, our method achieves around $\sim$5 dB and 147\% improvement in these two metrics, respectively. A high SSIM of 0.84 indicates that the restored image is almost the same as the ground truth. By contrast, the other three methods (i.e., BM3D, KSVD, and WNNM) show very limited performance in the quantitative comparison. Especially for the SSIM, all the three methods show almost no improvement.

The PAM images would provide a visual and qualitative comparison. Fig.~\ref{fig:4} shows two representative results from the testing set, where the drawbacks of the three traditional methods (BM3D, KSVD, and WNNM) can be identified. First, in some PAM images, including those recovered by BM3D (in the bottom row) and KSVD (in both rows), the noise cannot be effectively removed. The traditional methods rely on strict mathematical priors, which are associated with an accurate noise model and certain proper parameters. When the noise model is complex and/or certain parameters are not chosen properly, the de-noising performance would be highly degraded. In addition, the background regions of the PAM images recovered by the three traditional methods are found to be noisier (e.g., scattered noise observed in the cases of BM3D and KSVD) or not dark enough (e.g., a uniform red background observed in the case of WNNM (the full image in the bottom row), leading to low signal-to-background contrast) compared to those recovered by our CNN method. By contrast, our method is free from these drawbacks and thus produces the best de-noising results.

\begin{table*}[t]
	\caption{Quantitative Comparison Results Using the Synthetic Blood Vessel Dataset (Mean$\pm$Standard Deviation)}
	\small
	\begin{center}
		\begin{tabular}{c|ccccc}
			\hline
			Method & Noisy input & BM3D & KSVD & WNNM & Ours \\
			\hline
			PSNR (dB) & 23.14$\pm$5.88 & 24.75$\pm$5.82 & 23.44$\pm$5.54 & 25.17$\pm$6.03 & \bf{29.67$\pm$4.48} \\
			\hline
			SSIM & 0.37$\pm$0.23 & 0.47$\pm$0.21 & 0.36$\pm$0.18 & 0.50$\pm$0.22 & \bf{0.63$\pm$0.12} \\
			\hline
		\end{tabular}
	\end{center}
	\label{table:3}
\end{table*}

\begin{table*}[t]
	\caption{Quantitative Comparison Results Using the Real Noisy Blood Vessel Dataset (Mean$\pm$Standard Deviation)}
	\small
	\begin{center}
		\begin{tabular}{c|ccccc}
			\hline
			Method & Noisy input & BM3D & KSVD & WNNM & Ours \\
			\hline
			SNR (dB) & 26.64$\pm$5.03 & 36.54$\pm$2.50 & 27.53$\pm$4.37 & 42.33$\pm$6.03 & \bf{47.24$\pm$6.60} \\
			\hline
			CNR (dB) & 9.93$\pm$1.82 & 11.17$\pm$2.62 & 9.98$\pm$1.70 & 11.07$\pm$2.60 & \bf{12.37$\pm$2.66} \\
			\hline
		\end{tabular}
	\end{center}
	\label{table:4}
\end{table*}
\begin{figure*}[h!]
	\begin{center}
		\includegraphics[width=1\linewidth]{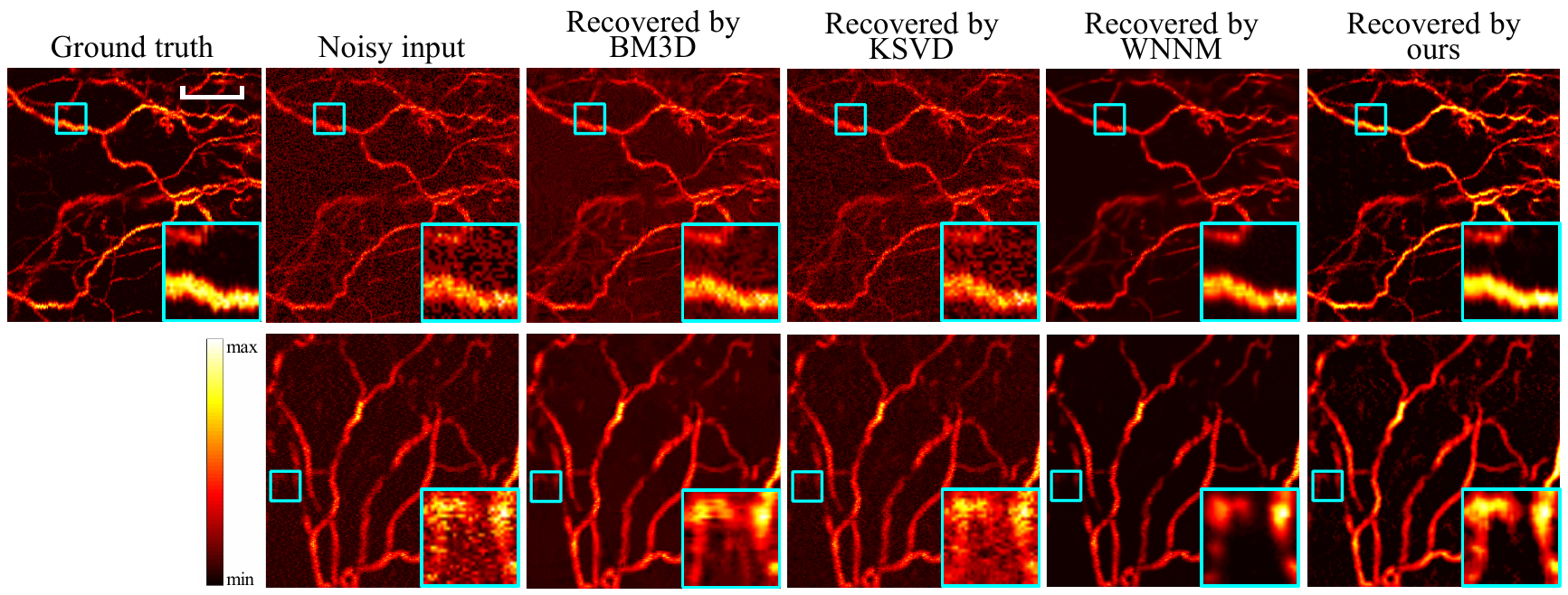}
	\end{center}
	\caption{Representative results of the mouse ear blood vessel dataset acquired by \emph{in vivo} experiment. Top raw: a representative sample from the synthetic noisy testing set; bottom row: a representative sample from the real noisy testing set. Scale bar: 250 \textmu m. All images, excluding zoom images, share the same scale bar. The values in the colorbar indicate relative PA intensity.}
	\label{fig:6}
\end{figure*}

\subsection{Real Noisy Leaf Vein Dataset for Validation}
\label{sec:real_leaf_results}

As mentioned in subsection~\ref{sec:noise_model}, the three types of noise added in PA A-line signals can well simulate the noise generated during PAM imaging. Therefore, we expect that the model trained on the synthetic dataset (subsection~\ref{sec:synthetic_results}) would also perform well for real noisy images.

As described earlier, 16 noisy leaf vein PAM images were experimentally acquired (subsection~\ref{sec:real_noisy_dataset}) using relatively low excitation light energy. The model trained on the synthetic dataset (subsection~\ref{sec:synthetic_results}) was used without further fine-tuning. Table~\ref{table:2} shows the quantitative results, and the no-reference metrics of SNR and CNR are used to compare the various de-noising methods. Similarly, two representative results of PAM images (out of 16) are shown in Fig.~\ref{fig:5}.

According to Table~\ref{table:2}, the proposed method also achieves the best quantitative results in terms of both SNR and CNR metrics. For SNR, compared to the noisy input, BM3D and KSVD produce little improvement ($<$8 dB), and WNNM enables appreciable improvement ($\sim$26 dB). As a comparison, our method results in the highest SNR improvement ($\sim$62 dB) when compared with the noisy input. As for CNR, compared to the noisy input, our method also brings about the highest improvement of $\sim$2.8 dB, while the other three methods have only slight improvements ($<$1 dB).

Similar to Fig.~\ref{fig:4}, for qualitative comparison, Fig.~\ref{fig:5} shows the de-noising results of the different methods. Overall, the images in Fig.~\ref{fig:5} are consistent with Table~\ref{table:2}. For example, our method enables the highest CNR (Table~\ref{table:2}), corresponding to the darkest background compared with the other methods (Fig.~\ref{fig:5}). Besides, similar to Fig.~\ref{fig:4}, our method renders the best de-noising performance while the other methods suffer from the drawbacks (i.e., being noisy and uniformly reddish background) mentioned previously.

\subsection{Blood Vessel Dataset: In Vivo Experiment}
\label{sec:in_vivo_results}
To further demonstrate that our method can be used for \emph{in vivo} imaging applications, de-noising of PAM images of mouse ear blood vessels was studied. Compared with leaf veins, blood vessels usually have relatively complex structure and low image quality, including (i) tortuous pattern, (ii) low SNR due to the laser safety limit in tissue, and (iii) artifacts (e.g., serrated boundaries) due to animal breathing and undesired absorbers in tissue. Therefore, it is expected to be more challenging to de-noise PAM images of blood vessels.

As mentioned in subsection~\ref{sec:dataset}, our \emph{in vivo} model was trained on the synthetic dataset (the clean blood vessel dataset (subsection~\ref{sec:clean_vessel_dataset}) and the corresponding synthetic noisy blood vessel dataset). Specifically, there are 1603 image pairs for training and 128 image pairs for testing. Similar to the synthetic leaf vein experiment, the model was first evaluated using the synthetic testing set (i.e., the 128 image pairs), and the results are shown in Table~\ref{table:3}. From Table~\ref{table:3}, the proposed method still outperforms other methods with noteworthy improvement for the \emph{in vivo} dataset.

As mentioned in subsection~\ref{sec:real_noisy_dataset}, 21 real noisy blood vessel images were experimentally acquired. Table~\ref{table:4} shows the quantitative results to compare the de-noising performance by different methods. Without ground truth, the no-reference metrics of SNR and CNR are used. The model trained on the synthetic blood vessel dataset was used here without additional transfer learning. According to Table~\ref{table:4}, our method still achieves the best de-noising performance in terms of quantitative metrics of SNR and CNR.

The de-noising performance can also be qualitatively observed in Fig.~\ref{fig:6}. The first row of Fig.~\ref{fig:6} shows the de-noising results of a representative sample from the synthetic noisy testing set, while the second row shows the de-noising results of a representative sample from the real noisy testing set. In general, the results in Fig.~\ref{fig:6} are consistent with Tables~\ref{table:3} and~\ref{table:4}. Our method produces clear blood vessels, while the other methods suffer from discontinuities (e.g., in the case of WNNM (the zoom image in the bottom row)) and severe interference by the background (e.g., in the cases of BM3D and KSVD). Although the model was only trained on the synthetic dataset, the noise in real \emph{in vivo} PAM images can still be effectively removed, which may be attributed to (i) the powerful image feature extraction and processing in the GAN-based model, and (ii) the inclusive noise model (i.e., three types of noise considered) for the training dataset.

\subsection{Self-Adaptability Demonstration}
\begin{figure}[h]
	\begin{center}
		\includegraphics[width=1\linewidth]{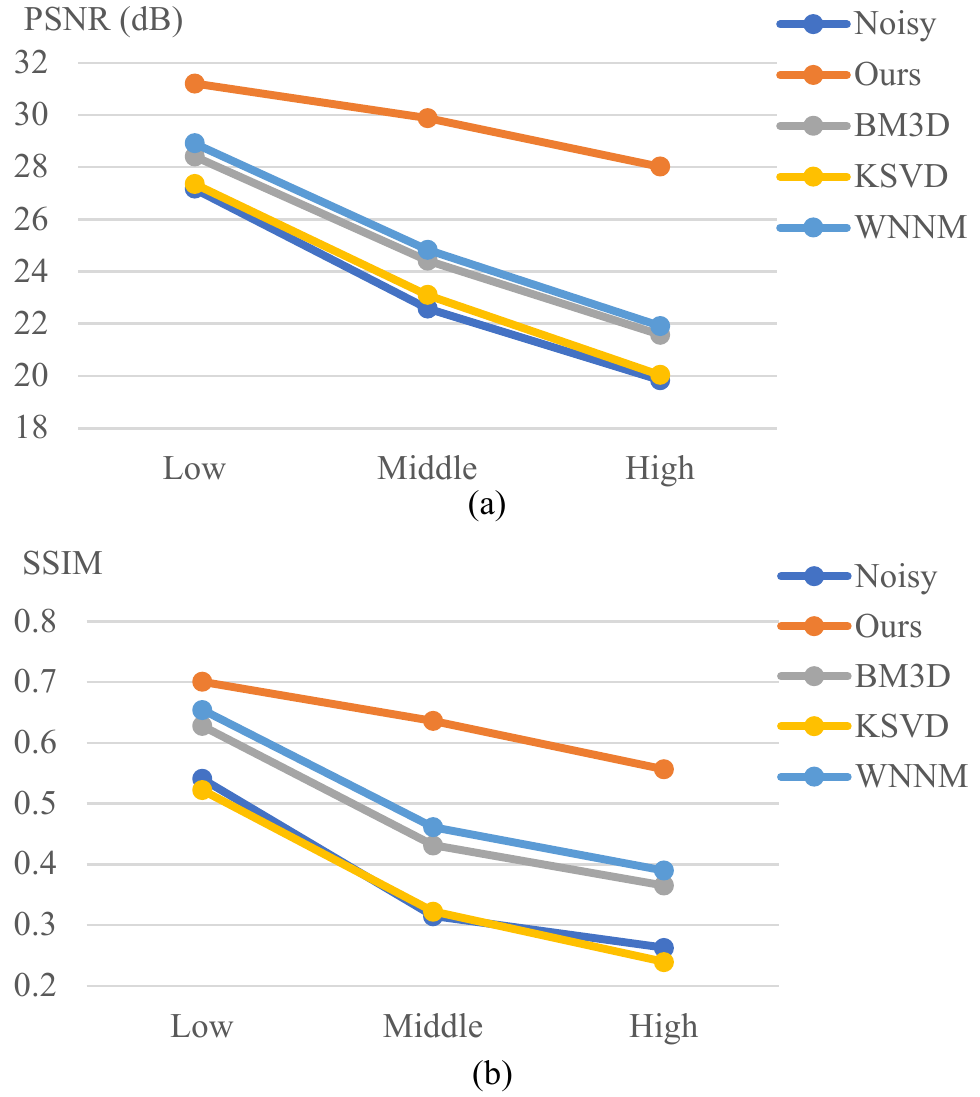}
	\end{center}
	\caption{Comparison of different de-noising methods applied to the noisy input images with three different noise levels (low noise level, middle noise level, and high noise level). The testing set of the synthetic \emph{in vivo} blood vessel dataset is used. (a) PSNR metric (average). (b) SSIM metric (average).}
	\label{fig:7}
\end{figure}

To illustrate the advantage of the self-adaptability of our method, a further comparison was demonstrated using the testing set of the synthetic \emph{in vivo} blood vessel dataset. Specifically, the testing set was split into three groups (low noise level, middle noise level, and high noise level) based on the noise level of the synthetic noisy images. Each group had a similar number of images. Different de-noising methods were applied to each group, respectively, for comparison. The results are shown in Fig.~\ref{fig:7}, where average PSNR and average SSIM are compared. Note that the three conventional methods adopt the manually set parameter to produce optimized results, while our method does not. It can be seen that our method achieves the highest PSNR and SSIM in all three groups. The margin of PSNR and SSIM between our method and the conventional methods becomes larger as the noise level of the noisy input image increases, which indicates that our method is more advantageous for de-noising images with high noise level. Therefore, the proposed deep learning-based method is promising for automatic processing, rather than manual or semi-manual processing, and would facilitate practical de-noising applications for PAM.

\begin{figure*}[h]
	\begin{center}
		\includegraphics[width=1\linewidth]{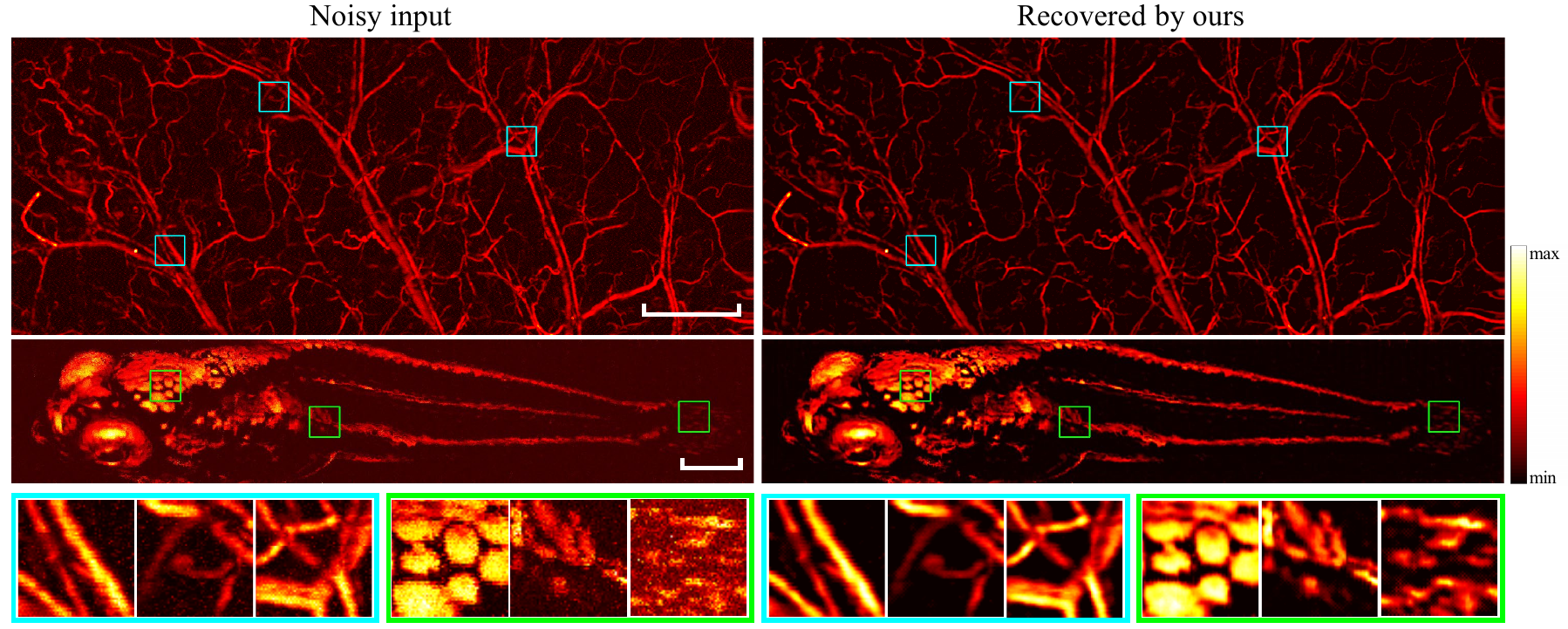}
	\end{center}
	\caption{Real noisy images with much larger size and with much different patterns. Top raw: mouse ear blood vessels (scale bar: 1 mm); middle row: zebrafish pigment (scale bars: 500 \textmu m); bottom row: representative zoom images of mouse ear blood vessels in the blue boxes (corresponding to the blue boxes from left to right in the top row) and zebrafish pigment in the green boxes (corresponding to the green boxes from left to right in the middle row). The values in the colorbar indicate relative PA intensity.}
	\label{fig:8}
\end{figure*}

In addition, we tested our method for real noisy images with much larger size and with much different patterns (zebrafish pigment vs. blood vessels). The results show that our method is robust. As shown in Fig.~\ref{fig:8}, without any transfer learning, our model trained on the synthetic blood vessel dataset can be used to de-noise wide-field PAM images. The two real noisy input images in Fig.~\ref{fig:8} are of $1500\times600$ pixels (corresponding physical size: $7.5 mm\times 3.0 mm$) for mouse ear blood vessels in the top row and $1448\times280$ pixels (corresponding physical size: $6.2 mm\times1.2 mm$) for zebrafish pigment in the bottom row. The pixel size in Fig.~\ref{fig:8} is much larger than that in the training set ($250\times250$ pixels). Furthermore, although the pattern of zebrafish pigment differs a lot from the pattern of blood vessels used for training (e.g., circle-like patterns in the zoom image around the zebrafish brain region), our model achieves excellent de-noising performance for zebrafish pigment, demonstrating the effectiveness of our comprehensive synthesis of three types of noise and the robust adaptability of our model to some extent.

\section{Discussion}
In this study, we proposed a PAM de-noising network with attention mechanism (GC block) and the combined loss. The proposed deep learning-based method demonstrated excellent de-noising performance compared to existing methods (BM3D, KSVD, and WNNM) on both phantom and \emph{in vivo} datasets. Specifically, quantitative metrics and qualitative observation of PAM images recovered by different methods were compared, and our method achieved the best results. As a self-comparison using the synthetic \emph{in vivo} blood vessel dataset, the model without GC blocks (while with other parts the same) led to the decreases of PSNR and SSIM by 0.59 dB and 0.04. The original U-Net without the modules (the GC blocks and the combined loss) resulted in the decreases of PSNR and SSIM by 2.85 and 0.04, showing the effectiveness of the modules in our proposed network. Besides, our network has other advantages, as elaborated in the following.

Conventional de-noising methods (e.g., BM3D, KSVD, and WNNM in this work) always require manually-set parameters to ensure the de-noising performance. For example, if the parameters of the assumed noise level are not carefully chosen, the de-noising performance could be greatly degraded. To ensure the optimal results using the conventional de-noising methods, it is often necessary to test different parameters of noise levels, and the best visual results need to be evaluated by humans. Note that in section~\ref{sec:results}, for a fair comparison, when applying the three conventional methods, the above de-noising approach was used for each image, and the optimized de-noising results were used. As expected, the procedure to find the best parameter of noise level is time-consuming due to the need for more computation time and manual evaluation. Further, for noisy images with different noise levels (e.g., due to different samples and different imaging system settings), the procedure to find the optimal parameters of noise level should be conducted for each image, which makes the whole process by using conventional algorithms prohibitively long. What is worse, the human evaluation makes the conventional algorithms inconvenient and impractical to use.

By contrast, our method has strength in self-adaptability. Once the training dataset is well prepared without serious data distribution errors and the training is effectively achieved, the trained model can be applied to images with a wide range of noise levels without the need for manual parameter setting. Therefore, in our demonstrations (the testing set of the synthetic noisy dataset (subsections~\ref{sec:synthetic_results} and~\ref{sec:in_vivo_results}) and the real noisy dataset (subsections~\ref{sec:real_leaf_results} and~\ref{sec:in_vivo_results})), our method does not require prior information of the noise level of the noisy input images and can still produce reasonable de-noised images without strange artifacts. This shows that our method can adapt to various noise levels.

Furthermore, the proposed method has another great advantage in much fast de-noising processing compared with the other methods. In our experiment, the three conventional methods (BM3D, KSVD, and WNNM) take from a few seconds to a few minutes to de-noise an input image with $256\times256$ pixels using a 3.20 GHz Intel Core i7-8700 CPU. By contrast, our method takes only 0.58 s to de-noise an input image of the same size using the same CPU. As CNNs are more suitable for GPU computing, our speed can be improved to 0.016 s/image using an Nvidia Titan RTX GPU. Hence, the proposed method is capable of real-time processing, which is essential for clinical PAM applications.

\section{Conclusion}
A deep learning-based de-noising method for PAM images was proposed and developed. The CNN effectively extracted multi-level image features and utilized attention mechanisms to improve the de-noising performance. The training schedule using GAN with the combined loss resulted in clean output. Three types of noise in PAM systems were considered when generating the synthetic noisy images for training and testing the model. Besides, real noisy images from both phantom and \emph{in vivo} datasets were used to validate the models trained on synthetic datasets. The results show that our proposed method performs well in de-noising PAM images by quantitative and qualitative comparison with other methods (BM3D, KSVD, and WNNM). By comparing PAM images recovered by different methods, our method has advantages in effective noise removal, well-suppressed background, and few artifacts. With the self-adaptability to various noise levels and the much fast speed for de-noising processing, the proposed method for PAM de-noising is promising for clinical applications.


\begin{thebibliography}{00}
	\bibitem{a1}
	L.~V. Wang and S.~Hu, ``Photoacoustic tomography: in vivo imaging from
	organelles to organs,'' \emph{science}, vol. 335, no. 6075, pp. 1458--1462,
	2012.
	
	\bibitem{a5}
	M.~Seong and S.-L. Chen, ``Recent advances toward clinical applications of
	photoacoustic microscopy: a review,'' \emph{Science China Life Sciences},
	vol.~63, no.~12, pp. 1798--1812, 2020.
	
	\bibitem{a7}
	D.~Wu, L.~Huang, M.~S. Jiang, and H.~Jiang, ``Contrast agents for photoacoustic
	and thermoacoustic imaging: a review,'' \emph{International journal of
		molecular sciences}, vol.~15, no.~12, pp. 23\,616--23\,639, 2014.
	
	\bibitem{a9}
	S.~Telenkov and A.~Mandelis, ``Signal-to-noise analysis of biomedical
	photoacoustic measurements in time and frequency domains,'' \emph{Review of
		Scientific Instruments}, vol.~81, no.~12, p. 124901, 2010.
	
	\bibitem{a10}
	B.~Stephanian, M.~T. Graham, H.~Hou, and M.~A.~L. Bell, ``Additive noise models
	for photoacoustic spatial coherence theory,'' \emph{Biomedical optics
		express}, vol.~9, no.~11, pp. 5566--5582, 2018.
	
	\bibitem{a11}
	S.~Mahmoodkalayeh, H.~Z. Jooya, A.~Hariri, Y.~Zhou, Q.~Xu, M.~A. Ansari, and
	M.~R. Avanaki, ``Low temperature-mediated enhancement of photoacoustic
	imaging depth,'' \emph{Scientific reports}, vol.~8, no.~1, pp. 1--9, 2018.
	
	\bibitem{a12}
	S.~H. Holan and J.~A. Viator, ``Automated wavelet denoising of photoacoustic
	signals for circulating melanoma cell detection and burn image
	reconstruction,'' \emph{Physics in Medicine \& Biology}, vol.~53, no.~12, p.
	N227, 2008.
	
	
	
	
	
	
	\bibitem{a18}
	C.~Lutzweiler, S.~Tzoumas, A.~Rosenthal, V.~Ntziachristos, and D.~Razansky,
	``High-throughput sparsity-based inversion scheme for optoacoustic
	tomography,'' \emph{IEEE transactions on medical imaging}, vol.~35, no.~2,
	pp. 674--684, 2015.
	
	\bibitem{a19}
	M.~Cao, T.~Feng, J.~Yuan, G.~Xu, X.~Wang, and P.~L. Carson, ``Spread spectrum
	photoacoustic tomography with image optimization,'' \emph{IEEE transactions
		on biomedical circuits and systems}, vol.~11, no.~2, pp. 411--419, 2016.
	
	\bibitem{a20}
	E.~R. Hill, W.~Xia, M.~J. Clarkson, and A.~E. Desjardins, ``Identification and
	removal of laser-induced noise in photoacoustic imaging using singular value
	decomposition,'' \emph{Biomedical optics express}, vol.~8, no.~1, pp. 68--77,
	2017.
	
	\bibitem{a21}
	I.~U. Haq, R.~Nagaoka, S.~Siregar, and Y.~Saijo, ``Sparse-representation-based
	denoising of photoacoustic images,'' \emph{Biomedical Physics \& Engineering
		Express}, vol.~3, no.~4, p. 045014, 2017.
	
	\bibitem{a22}
	M.~Zhou, H.~Xia, H.~Lan, T.~Duan, H.~Zhong, and F.~Gao, ``Wavelet de-noising
	method with adaptive threshold selection for photoacoustic tomography,'' in
	\emph{2018 40th Annual International Conference of the IEEE Engineering in
		Medicine and Biology Society (EMBC)}.\hskip 1em plus 0.5em minus 0.4em\relax
	IEEE, 2018, pp. 4796--4799.
	
	
	\bibitem{a24}
	S.~Siregar, R.~Nagaoka, I.~U. Haq, and Y.~Saijo, ``Non local means denoising in
	photoacoustic imaging,'' \emph{Japanese Journal of Applied Physics}, vol.~57,
	no. 7S1, p. 07LB06, 2018.
	
	
	\bibitem{a26}
	M.~Zhou, H.~Zhao, H.~Xia, J.~Zhang, Z.~Liu, C.~Liu, and F.~Gao, ``De-noising of
	photoacoustic sensing and imaging based on combined empirical mode
	decomposition and independent component analysis,'' \emph{Journal of
		biophotonics}, vol.~12, no.~8, p. e201900042, 2019.
	
	\bibitem{a27}
	M.~Zhou, H.~Xia, H.~Zhong, J.~Zhang, and F.~Gao, ``A noise reduction method for
	photoacoustic imaging in vivo based on emd and conditional mutual
	information,'' \emph{IEEE Photonics Journal}, vol.~11, no.~1, pp. 1--10,
	2019.
	
	\bibitem{a28}
	E.~Najafzadeh, P.~Farnia, S.~N. Lavasani, M.~Basij, Y.~Yan, H.~Ghadiri,
	A.~Ahmadian, and M.~Mehrmohammadi, ``Photoacoustic image improvement based on
	a combination of sparse coding and filtering,'' \emph{Journal of biomedical
		optics}, vol.~25, no.~10, p. 106001, 2020.
	
	
	\bibitem{a30}
	A.~Hariri, K.~Alipour, Y.~Mantri, J.~P. Schulze, and J.~V. Jokerst, ``Deep
	learning improves contrast in low-fluence photoacoustic imaging,''
	\emph{Biomedical optics express}, vol.~11, no.~6, pp. 3360--3373, 2020.
	
	\bibitem{a31}
	K.~Suzuki, ``Overview of deep learning in medical imaging,'' \emph{Radiological
		physics and technology}, vol.~10, no.~3, pp. 257--273, 2017.
	
	\bibitem{a32}
	D.~Allman, A.~Reiter, and M.~A.~L. Bell, ``Photoacoustic source detection and
	reflection artifact removal enabled by deep learning,'' \emph{IEEE
		transactions on medical imaging}, vol.~37, no.~6, pp. 1464--1477, 2018.
	
	\bibitem{a33}
	N.~Davoudi, X.~L. De{\'a}n-Ben, and D.~Razansky, ``Deep learning optoacoustic
	tomography with sparse data,'' \emph{Nature Machine Intelligence}, vol.~1,
	no.~10, pp. 453--460, 2019.
	
	\bibitem{a34}
	J.~Zhou, D.~He, X.~Shang, Z.~Guo, S.-L. Chen, and J.~Luo, ``Photoacoustic
	microscopy with sparse data by convolutional neural networks,''
	\emph{Photoacoustics}, vol.~22, p. 100242, 2021.
	
	\bibitem{a38}
	J.~Chen, J.~Chen, H.~Chao, and M.~Yang, ``Image blind denoising with generative
	adversarial network based noise modeling,'' in \emph{Proceedings of the IEEE
		Conference on Computer Vision and Pattern Recognition}, 2018, pp. 3155--3164.
	
	\bibitem{a40}
	K.~Lin, T.~H. Li, S.~Liu, and G.~Li, ``Real photographs denoising with noise
	domain adaptation and attentive generative adversarial network,'' in
	\emph{Proceedings of the IEEE/CVF Conference on Computer Vision and Pattern
		Recognition Workshops}, 2019, pp. 0--0.
	
	\bibitem{a42}
	Q.~Yang, P.~Yan, Y.~Zhang, H.~Yu, Y.~Shi, X.~Mou, M.~K. Kalra, Y.~Zhang,
	L.~Sun, and G.~Wang, ``Low-dose ct image denoising using a generative
	adversarial network with wasserstein distance and perceptual loss,''
	\emph{IEEE transactions on medical imaging}, vol.~37, no.~6, pp. 1348--1357,
	2018.
	
	\bibitem{a43}
	D.~Mishra, S.~Chaudhury, M.~Sarkar, and A.~S. Soin, ``Ultrasound image
	enhancement using structure oriented adversarial network,'' \emph{IEEE Signal
		Processing Letters}, vol.~25, no.~9, pp. 1349--1353, 2018.
	
	\bibitem{a44}
	K.~J. Halupka, B.~J. Antony, M.~H. Lee, K.~A. Lucy, R.~S. Rai, H.~Ishikawa,
	G.~Wollstein, J.~S. Schuman, and R.~Garnavi, ``Retinal optical coherence
	tomography image enhancement via deep learning,'' \emph{Biomedical optics
		express}, vol.~9, no.~12, pp. 6205--6221, 2018.
	
	\bibitem{a}
	A.~Sharma and M.~Pramanik, ``Convolutional neural network for resolution
	enhancement and noise reduction in acoustic resolution photoacoustic
	microscopy,'' \emph{Biomedical Optics Express}, vol.~11, no.~12, pp.
	6826--6839, 2020.
	
	\bibitem{x2}
	H.~Zhao, Z.~Ke, F.~Yang, K.~Li, N.~Chen, L.~Song, C.~Zheng, D.~Liang, and
	C.~Liu, ``Deep learning enables superior photoacoustic imaging at ultralow
	laser dosages,'' \emph{Advanced Science}, vol.~8, no.~3, p. 2003097, 2021.
	
	\bibitem{a47}
	K.~Dabov, A.~Foi, V.~Katkovnik, and K.~Egiazarian, ``Image denoising by sparse
	3-d transform-domain collaborative filtering,'' \emph{IEEE Transactions on
		image processing}, vol.~16, no.~8, pp. 2080--2095, 2007.
	
	\bibitem{a48}
	S.~Gu, L.~Zhang, W.~Zuo, and X.~Feng, ``Weighted nuclear norm minimization with
	application to image denoising,'' in \emph{Proceedings of the IEEE conference
		on computer vision and pattern recognition}, 2014, pp. 2862--2869.
	
	\bibitem{a46}
	I.~Goodfellow, J.~Pouget-Abadie, M.~Mirza, B.~Xu, D.~Warde-Farley, S.~Ozair,
	A.~Courville, and Y.~Bengio, ``Generative adversarial nets,'' \emph{Advances
		in neural information processing systems}, vol.~27, 2014.
	
	\bibitem{b}
	P.~Patidar, M.~Gupta, S.~Srivastava, and A.~K. Nagawat, ``Image de-noising by
	various filters for different noise,'' \emph{International journal of
		computer applications}, vol.~9, no.~4, pp. 45--50, 2010.
	
	\bibitem{c}
	B.~Cohen \emph{et~al.}, ``New maximum likelihood motion estimation schemes for
	noisy ultrasound images,'' \emph{Pattern Recognition}, vol.~35, no.~2, pp.
	455--463, 2002.
	
	\bibitem{a50}
	Y.~Cao, J.~Xu, S.~Lin, F.~Wei, and H.~Hu, ``Gcnet: Non-local networks meet
	squeeze-excitation networks and beyond,'' in \emph{Proceedings of the
		IEEE/CVF International Conference on Computer Vision Workshops}, 2019, pp.
	0--0.
	
	\bibitem{a55}
	D.~Ulyanov, A.~Vedaldi, and V.~Lempitsky, ``Instance normalization: The missing
	ingredient for fast stylization,'' \emph{arXiv preprint arXiv:1607.08022},
	2016.
	
	\bibitem{a56}
	J.~Johnson, A.~Alahi, and L.~Fei-Fei, ``Perceptual losses for real-time style
	transfer and super-resolution,'' in \emph{European conference on computer
		vision}.\hskip 1em plus 0.5em minus 0.4em\relax Springer, 2016, pp. 694--711.
	
	\bibitem{a57}
	S.~Ren, K.~He, R.~Girshick, and J.~Sun, ``Faster r-cnn: Towards real-time
	object detection with region proposal networks,'' \emph{Advances in neural
		information processing systems}, vol.~28, pp. 91--99, 2015.
	
	\bibitem{a58}
	K.~Simonyan and A.~Zisserman, ``Very deep convolutional networks for
	large-scale image recognition,'' \emph{arXiv preprint arXiv:1409.1556}, 2014.
	
	\bibitem{x1}
	Z.~Guo, Z.~Ye, W.~Shao, L.~Jing, and S.-L. Chen, ``Miniature probe for
	optomechanical focus-adjustable optical-resolution photoacoustic endoscopy,''
	\emph{arXiv preprint arXiv:1911.09522}, 2019.
	
	\bibitem{a60}
	D.~P. Kingma and J.~Ba, ``Adam: A method for stochastic optimization,''
	\emph{arXiv preprint arXiv:1412.6980}, 2014.
	
	\bibitem{a61}
	M.~Aharon, M.~Elad, and A.~Bruckstein, ``K-svd: An algorithm for designing
	overcomplete dictionaries for sparse representation,'' \emph{IEEE
		Transactions on signal processing}, vol.~54, no.~11, pp. 4311--4322, 2006.
	
\end{thebibliography}
\end{document}